\def\OSML{\textsc{Osml}\xspace}
\def\HTZT{\textsc{HtZt}\xspace}
\def\MONO{\textsc{Monotop}\xspace}

\documentclass[PAPER, atlasdraft=false, cernpreprint, USenglish, texmf, orcidlogo]{atlasdoc}
\usepackage{atlaspackage}
\usepackage{atlasxref}
\usepackage{atlasjetetmiss}
\usepackage{atlassnippets}
\usepackage{atlasmisc}
\usepackage{atlasbiblatex}

\usepackage{atlasphysics}
\graphicspath{{logos/}{figures/}}

\usepackage{ANA-EXOT-2021-02-PAPER-defs}
\usepackage{multirow}
\usepackage{comment}
\usepackage{amsmath}
\usepackage{slashed}
\usepackage{ulem}
\usepackage{tabularx}
\usepackage{booktabs}

\DeclareUnicodeCharacter{2212}{-}

\AtlasTitle{Combination of searches for singly produced vector-like top quarks in \textit{pp} collisions at $\sqrt{s} = 13$~\TeV with the ATLAS detector}

\AtlasAbstract{%
A combination of searches for the single production of
vector-like top quarks ($T$) is presented. These analyses are based on proton--proton collisions
at $\sqrt{s}=13~\tev$ recorded in 2015--2018 with the ATLAS detector at the Large Hadron Collider, corresponding
to an integrated luminosity
of 139~fb$^{-1}$.  The \VLTa decay modes considered in this combination are into a top quark and
either a Standard Model Higgs boson or a $Z$ boson ($T \to Ht$ and $T \to Zt$).
The individual searches used in the combination are differentiated by the number of leptons ($e$, $\mu$)
in the final state. The observed data are found to be in good agreement with the Standard Model background prediction.
Interpretations are provided for a range of masses and couplings of
the vector-like top quark for benchmark models and generalized representations
in terms of 95\% confidence level limits. For a benchmark signal
prediction of a vector-like top quark \SU2 singlet with electroweak coupling, $\kappa$, of 0.5,
masses below 2.1~TeV are excluded, resulting in the most restrictive limits to date.
}

\AtlasRefCode{EXOT-2021-02}

\PreprintIdNumber{CERN-EP-2024-203}

\AtlasJournalRef{\PRD 111 (2025) 012012}
\AtlasDOI{10.1103/PhysRevD.111.012012}

\AtlasCoverEgroupAnalysisTeam{atlas-exot-2021-02-analysis-team@cern.ch}

\hypersetup{pdftitle={ATLAS document},pdfauthor={The ATLAS Collaboration}}

\begin{document}

\maketitle

\tableofcontents

%
%

%
%

% The next lines are included from the .//Sections/Introduction.tex input file
\section{Introduction}
\label{sec:introduction}

The formulation of electroweak interactions arising from a spontaneously broken gauge symmetry is a cornerstone of the Standard Model (SM).  Experiments over the past four decades have confirmed this hypothesis, most notably through the precision measurements of the LEP and SLC collider programs~\cite{PEWT1,PEWT2}. A major milestone was achieved when the ATLAS and CMS Collaborations reported the discovery~\cite{HiggsATLAS,HiggsCMS} of a new particle produced at CERN's Large Hadron Collider (LHC) possessing properties consistent with those predicted for the SM Higgs boson ($H$). The electroweak symmetry-breaking mechanism, where a weak-isospin doublet of fundamental scalar fields obtains a vacuum expectation value, remains the simplest hypothesis.

Following the discovery of the Higgs boson, the SM still cannot be considered a complete description of nature.  For example, the theory does not explain the number of fermion generations and their mass hierarchy and mixing angles, nor the origin of the matter--antimatter asymmetry in the universe.  It also does not have a viable dark matter particle.  Therefore, the SM is generally regarded as a low-energy approximation of a more fundamental theory with new degrees of freedom and symmetries that would become manifest at higher energy. In fact, the SM violates a concept of naturalness~\cite{Naturalness} when extrapolated to energies above the electroweak scale. When extrapolated to the energy scale of new physics, a fine-tuning of the theory is required. The fine-tuning can be mitigated by the introduction of new interactions that cancel out the quadratic divergences in the Higgs boson mass.
To this effect, several explanations are proposed in theories beyond the SM (BSM).

Proposed new models typically address the naturalness problem by postulating a new symmetry.  For example, supersymmetry is a Bose--Fermi symmetry, and the new states related to the SM bosons and fermions by this symmetry introduce new interactions that cancel out the quadratically divergent ones~\cite{Golfand:1971iw,Volkov:1973ix,Wess:1974tw,Wess:1974jb,Ferrara:1974pu,Salam:1974ig}.  Alternatively, the symmetry could be a spontaneously broken global symmetry of the extended theory, with the Higgs boson emerging as a pseudo-Nambu--Goldstone boson~\cite{StrongEWSB}.  Examples of models that implement this idea are Little Higgs~\cite{LittleHiggs,LittleHiggsRev} and Composite Higgs~\cite{CompHiggs1,CompHiggs2} models.  The new states realizing the enhanced symmetry are generically strongly coupled resonances of new confining dynamics.  These include vector-like quarks (VLQs), defined as color-triplet spin-$1/2$ fermions whose left- and right-handed chiral components have the same transformation properties under the weak-isospin gauge group.  Such quarks could mix with like-charge SM quarks~\cite{VLQmixing,delAguila}, and the mixing of the SM top quark with a charge $+2e/3$ vector-like quark (where $e$ is the electric charge of the electron) could play a role in regulating the sensitivity to the Higgs boson mass. Hence, VLQs emerge as a characteristic feature of several non-supersymmetric models~\cite{FermNatural}.

In order for VLQs to mix with the SM quarks~\cite{VLQmixing,delAguila}, some constraints are needed on their allowed quantum numbers to preserve gauge invariance. Therefore, in order to generate Yukawa terms in the Lagrangian without changing the scalar sector, only seven renormalizable possibilities~\cite{delAguila:2000rc} are allowed, which are summarized in Table~\ref{tab:multiplets}. The vector-like $T$ and $B$ have electric charge $+2e/3$ and $-1e/3$, respectively. VLQs with other electric charges could also exist, such as the $X$ and $Y$ quarks with $+5e/3$ and $-4e/3$, respectively.

\begin{table}[tb]
\begin{center}\vspace{0.3cm}
\caption[Possible multiplets of vector-like]{Possible VLQ multiplets allowed to mix with the SM quarks.}
\begin{tabular}{l|cccccccc}\toprule
\textbf{VLQ} & $(T)$ & $(B)$ & $\begin{pmatrix}
T \\ B
\end{pmatrix}$ & $\begin{pmatrix}
X \\ T
\end{pmatrix}$ & $\begin{pmatrix}
B \\ Y
\end{pmatrix}$ & $\begin{pmatrix}
X \\ T \\ B
\end{pmatrix}$ & $\begin{pmatrix}
T \\ B \\ Y
\end{pmatrix}$ \\[5pt]
\textbf{Isospin} & 0 & 0 &1/2 & 1/2 &1/2 & 1 & 1\\
\textbf{Hypercharge} & $+2/3$ & $-1/3$ & $+1/6$ & $+7/6$ & $-5/6$ & $+2/3$ & $-1/3$ \\ \bottomrule
\end{tabular}
\label{tab:multiplets}
\vspace{-0.5cm}
\end{center}
\end{table}

The renormalizable electroweak representations consist of $(T)$ or $(B)$ \SU2 singlets, $(X,T)$, $(T,B)$, or $(B,Y)$ doublets
and $(X,T,B)$ or $(T,B,Y)$ triplets. In all representations, they couple to the SM quarks via an exchange of charged $(W^\pm)$ or neutral $(Z,H)$ bosons. The  simplified Lagrangian~\cite{Fuks2017} below summarizes the interaction of the VLQ and SM quarks:

\begin{equation}\label{eqn:SimpLag}
\mathcal{L} = \sum_{Q, q, \zeta} \left[ \frac{g_w}{\sqrt{2}}{\kappa}_{\zeta}^{Qq}       \bar{Q}\slashed{W} P_{\zeta} q
+ \frac{g_w}{2c_W} \tilde{\kappa}_{\zeta}^{Qq} \bar{Q}\slashed{Z} P_{\zeta} q
+  \hat{\kappa}_{\zeta}^{Qq} H \bar{Q} P_{\zeta} q\right] + \mathrm{H.c.,}
\end{equation}

where $Q$ represents a VLQ, $\zeta$  represents the chirality with $P_\zeta$ being the corresponding projection operator, $q$ represents a SM quark of up or down type, and the
electroweak couplings ${\kappa}_\zeta^{Qq}, \tilde{\kappa}_\zeta^{Qq},$ and $\hat{\kappa}_\zeta^{Qq}$ determine
the coupling strengths between $Q$ and $q$ when mediated by the $W,Z,$ and $H$ bosons, respectively.
The mass hierarchy of the SM quarks suggests that VLQs interact predominantly with the third-generation SM quarks~\cite{Okada,JA_new}.
Hence, VLQ interactions with lighter generations are set to zero in the
simplified representation of Eq.~(\ref{eqn:SimpLag}). Furthermore, this
formulation assumes there are no additional mediators other than the  $W,Z,$ and
$H$  bosons~\cite{HeavyVector}.

\begin{figure*}[tb]
\centering
\includegraphics{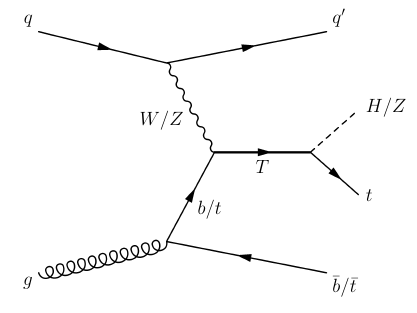}
\caption{Illustrative leading-order Feynman diagram of single vector-like $T$ production in association with a $t$ or $b$ quark and subsequent decay into either $Ht$ or $Zt$.}
\label{fig:singleT}
\end{figure*}

In proton--proton ($pp$) collisions, VLQs can be produced singly via the electroweak interaction (as illustrated by the leading-order (LO) Feynman diagram in Figure~\ref{fig:singleT}) or in pairs via the strong interaction. While the cross-section for pair production is generally given by quantum chromodynamics, the single-production cross-section explicitly depends on the coupling of the VLQs to SM electroweak bosons. Search strategies for VLQs were outlined previously~\cite{ContinoServant,JA_TP,TP_guide,JA_new,Matsedonskyi:2014mna}.  Results of searches for chiral fourth-generation quarks apply, though interpreting the exclusions was difficult when the quarks were assumed to decay entirely via the charged-current process. When VLQs are added to the SM, flavor-changing tree-level neutral-current decays of such new heavy quarks appear~\cite{delAguila2}, while they are not present for SM quarks at tree level and are also highly suppressed at loop level by the Glashow–Iliopoulos–Maiani (GIM) mechanism~\cite{GIM}. Following Ref.~\cite{Roy:2020fqf}, the relative couplings of VLQs to $W, Z$ and $H$ bosons are given in terms of the parameters $\xi_W,$ $\xi_Z$ and $\xi_H$, respectively. In the asymptotic limit of large VLQ mass, the $\xi$ parameters correspond to the branching ratios of the $T$ quark decay to $W, Z$ and $H$ bosons. The asymptotic limit holds to good approximation for VLQ masses above 1~TeV.  In the $T$ \SU2 singlet representation, $\xi_W$ = 0.5 and $\xi_Z = \xi_H = 0.25$. If the $T$ quarks are part of an $(X, T)$ or $(T, B)$ \SU2 doublet then  $\xi_W$ = 0.0, and $\xi_Z = \xi_H= 0.5$.

Vector-like $T$ and $B$ quarks were searched for both at ATLAS and CMS in Run 1, and more recently with the Run 2 data. The Run 2 searches focused on VLQ pair production~\cite{combPaper,osmlPaper2018,ATLASfullHad,ATLASssb,ATLASfullHad,ATLASpairWtX,ATLASpairHtX,ATLASpairWbX,ATLASpairLepMet,CMSppAllhad,CMSppOsml,CMSppleptons,CMSppWbWb,EXOT-2018-58,CMS-B2G-20-011, EXOT-2019-06, EXOT-2019-08}, typically setting limits on narrow-width resonances, and recently also on VLQ single production~\cite{osmlPaper2018,EXOT-2022-40,ATLASsingleTWb,ATLASssb,CMSspAllHad,CMSspWb,CMSspHb, EXOT-2016-35, CMS-B2G-17-007, EXOT-2019-07, OSMLPaper2024, EXOT-2018-52, EXOT-2022-40, CMS-B2G-19-004, CMS-B2G-21-007, CMS-B2G-19-001}. Searches for VLQ are summarized by the ATLAS and CMS Collaborations in Refs.~\cite{EXOT-2023-14, CMS-EXO-23-006}. The most stringent limits on $T$- and $B$-quark masses come from the latest ATLAS and CMS searches. The excluded masses for VLQs depend on the branching ratio.
In the case of \SU2 doublet representations $(T, B)$ where both VLQs are considered and assumed to be mass degenerate, $T$ quark masses below $1.6\, \TeV$~\cite{EXOT-2019-08} are excluded at 95\% confidence level (CL). When no assumption is made on the branching ratio or mass degeneracy, then $T$ quarks with masses below $1.5\,\TeV$~\cite{CMS-B2G-20-011} are excluded. In the case of \SU2 singlet representations masses below $1.5\,\TeV$~\cite{CMS-B2G-20-011} are excluded. The pair production limits assume a narrow-width resonance, an assumption not made for single production. In single production, for which the interpretation of results is more challenging due to the additional coupling factor, analyses have set limits in the coupling/mixing-angle-vs.-mass plane, reaching up to nearly $2~\tev$~\cite{OSMLPaper2024} for large coupling and mixing angles.

This paper describes a combination of three analyses, differentiated by the number of light leptons ($e$, $\mu$) in the final state. The search regions for each analysis, including the control regions, are designed to be orthogonal as they target different decay channels. This is achieved by requiring exactly zero leptons (electrons or muons) in the ``\MONO'' analysis~\cite{EXOT-2022-40}, exactly one lepton in the ``\HTZT'' analysis~\cite{HtZtPaper2023}, and two or more leptons in the ``\OSML'' (Opposite-Sign Multi-Lepton) analysis~\cite{OSMLPaper2024}. %
All searches use the full Run~2 ATLAS dataset collected with the ATLAS detector during 2015--2018, corresponding to 139~fb$^{-1}$ of $pp$ collisions at $\sqrt{s}=13$~TeV.
The searches are combined taking into account correlations in the background modeling and systematic uncertainties. In the absence of a significant excess above the SM expectation,
the results are used to set upper limits on the single production of $T$ quarks for several scenarios of the mass, the universal coupling strength $\kappa$, and the relative couplings to $W$, $Z$ and Higgs bosons.
This is the first combination of searches for single \VLTa production by the ATLAS Collaboration and the results provide the most restrictive bounds to date.

% End of text imported from the .//Sections/Introduction.tex input file

% The next lines are included from the .//Sections/Detector.tex input file
\section{ATLAS detector}
\label{sec:detector}

\newcommand{\AtlasCoordFootnote}{%
ATLAS uses a right-handed coordinate system with its origin at the nominal interaction point (IP)
in the center of the detector and the \(z\)-axis along the beam pipe.
The \(x\)-axis points from the IP to the center of the LHC ring,
and the \(y\)-axis points upwards.
Polar coordinates \((r,\phi)\) are used in the transverse plane,
\(\phi\) being the azimuthal angle around the \(z\)-axis.
The pseudorapidity is defined in terms of the polar angle \(\theta\) as \(\eta = -\ln \tan(\theta/2)\) and is equal to the rapidity
$ y = \frac{1}{2} \ln \left( \frac{E + p_z c}{E - p_z c} \right) $ in the relativistic limit.
Angular distance is measured in units of \(\Delta R \equiv \sqrt{(\Delta y)^{2} + (\Delta\phi)^{2}}\).}

The ATLAS experiment~\cite{PERF-2007-01} at the LHC is a multipurpose particle detector
with a forward--backward symmetric cylindrical geometry and a near \(4\pi\) coverage in
solid angle.\footnote{\AtlasCoordFootnote}
It consists of an inner tracking detector surrounded by a thin superconducting solenoid
providing a \qty{2}{\tesla} axial magnetic field, electromagnetic and hadronic calorimeters, and a muon spectrometer.
The inner tracking detector covers the pseudorapidity range \(|\eta| < 2.5\).
It consists of silicon pixel, silicon microstrip, and transition radiation tracking detectors.
Lead/liquid-argon (LAr) sampling calorimeters provide electromagnetic (EM) energy measurements
with high granularity within the region \(|\eta|< 3.2\).
A steel/scintillator-tile hadronic calorimeter covers the central pseudorapidity range (\(|\eta| < 1.7\)).
The endcap and forward regions are instrumented with LAr calorimeters
for EM and hadronic energy measurements up to \(|\eta| = 4.9\).
The muon spectrometer surrounds the calorimeters and is based on
three large superconducting air-core toroidal magnets with eight coils each.
The field integral of the toroids ranges between \num{2.0} and \qty{6.0}{\tesla\metre}
across most of the detector.
The muon spectrometer includes a system of precision tracking chambers up to \(|\eta| = 2.7\) and fast detectors for triggering up to \(|\eta| = 2.4\).
The luminosity is measured mainly by the LUCID--2~\cite{LUCID2} detector, which is located close to the beampipe.
A two-level trigger system is used to select events~\cite{TRIG-2016-01}.
The first-level trigger is implemented in hardware and uses a subset of the detector information
to accept events at a rate below \qty{100}{\kHz}.
This is followed by a software-based trigger that
reduces the accepted event rate to \qty{1}{\kHz} on average
depending on the data-taking conditions.
A software suite~\cite{SOFT-2022-02} is used in data simulation, in the reconstruction
and analysis of real and simulated data, in detector operations, and in the trigger and data acquisition
systems of the experiment.

% End of text imported from the .//Sections/Detector.tex input file

%
%
%

% The next lines are included from the .//Sections/DataAndSimulation.tex input file
\section{Data and simulated event samples}
\label{sec:datatrig}

This combination uses $pp$ collision data collected with the ATLAS detector during the 2015--2018 data
taking period, at a collision energy of $\sqrt{s}=13~\tev$, corresponding to an integrated
luminosity of $139~\ifb$~\cite{DAPR-2021-01,LUCID2}.
Only events recorded with a single-electron trigger, a single-muon trigger or a \met\ trigger~\cite{TRIG-2018-05, TRIG-2018-01, TRIG-2019-02}
under stable beam conditions, and for which all relevant detector subsystems were operational~\cite{DAPR-2018-01}, are used in
these analyses. The trigger requirements of the individual analyses can be found
in the respective publications~\cite{EXOT-2022-40,HtZtPaper2023,OSMLPaper2024}.

Monte Carlo (MC) simulated events are used to evaluate background contamination and modeling, signal acceptance,
optimization of the event selection, and  evaluation of systematic uncertainties. They are centrally produced with the well-established
ATLAS event generation procedure~\cite{Aad:2010ah}. For all samples, a full simulation of the ATLAS detector was performed
using \GEANT4~\cite{Agostinelli:2002hh}. A faster simulation, where the full \GEANT4 simulation of the calorimeter response
is replaced by a detailed parameterization of the shower shapes~\cite{ATL-PHYS-PUB-2010-013}, was adopted for some of the
samples used to estimate systematic uncertainties.

The effect of multiple interactions in the same and neighboring bunch crossings (pile-up) was modeled by overlaying
simulated hard-scattering events with inelastic $pp$ events generated with \textsc{Pythia8} (v8.186)~\cite{Sjostrand:2007gs}
using the NNPDF2.3LO parton distribution function (PDF) set~\cite{Ball:2012cx} and the A3 set of tuned parameters (tune)~\cite{ATL-PHYS-PUB-2016-017}.
The MC events were weighted to reproduce the distribution of the average number of interactions per bunch crossing
($\left<\mu \right>$) observed in the data.

Table~\ref{tab:mcsamples} summarizes the setups used in the simulated SM background samples. %
Except for the samples generated with \SHERPA~\cite{Gleisberg:2008ta}, all samples are interfaced with \PYTHIAV{8.2}~\cite{Sjostrand:2014zea} for the modeling of the parton shower (PS),
hadronization and underlying event (UE) using the NNPDF2.3LO PDF set~\cite{Ball:2014uwa} and the A14 tune~\cite{ATL-PHYS-PUB-2014-021}.
Alternative samples to study the impact of systematic uncertainties due to the PS and hadronization model
are instead interfaced with \HERWIGV{7}~\cite{Bellm:2015jjp,Bahr:2008pv}
using the MMHT2014LO PDF set~\cite{Harland-Lang:2014zoa} and the H7UE tune~\cite{Bellm:2015jjp}. %
Additional details and other uncertainties in background modeling are given in the respective references~\cite{EXOT-2022-40,HtZtPaper2023,OSMLPaper2024}.
The \EvtGen\ 1.6.0~\cite{Lange:2001uf} program was used to model $b$- and $c$-hadron decays
for all samples showered using \PYTHIAV{8.2} or \HERWIGV{7}, with the
exception of the $t\bar{t}V$ ($V=W,Z$) samples, where \EvtGen\ 1.2.0 is used instead.

\begin{table}[tb]
\centering
\scriptsize
\caption{Overview of the nominal simulated background samples, including information about the matrix element (ME) generator and parton distribution function (PDF) set, parton shower (PS), set of tuned parameters (Tune), and higher-order cross-section used for normalization.
For single top the $t$- and $s$-channel production cross-sections are calculated at next-to-NLO (NNLO) in QCD, while the $Wt$ production cross-section is calculated at NLO and includes third-order corrections of soft-gluon emissions by resuming next-to-next-to-leading logarithm (NNLL) terms.
}
\begin{tabular}{llllll}
\toprule
Process & ME event generator   & ME PDF & PS and  & Tune & Cross-section   \\
&                                       &              & hadronization &         & calculation   \\
\hline
$W/Z$+jets  & \SHERPA~2.2.1~\cite{Gleisberg:2008ta}         & NNPDF3.0NNLO~\cite{Ball:2014uwa}  & \SHERPA ~\cite{Gleisberg:2008ta}       & Default & NNLO~\cite{Catani:2009sm}    \\
$t\bar t W/Z$        & \MGMCatNLOV 2.3.3~\cite{Alwall:2014hca}       & NNPDF3.0NLO~\cite{Ball:2014uwa}   & \PYTHIAV{8.2}~\cite{Sjostrand:2014zea}    & A14     & NLO    \\
$tZ$               & \MGMCatNLOV 2.3.3~\cite{Alwall:2014hca}       & NNPDF2.3LO~\cite{Ball:2012cx}     & \PYTHIAV{8.2}~\cite{Sjostrand:2014zea}    & A14     & NLO    \\
$tWZ$              & \MGMCatNLOV 2.3.3~\cite{Alwall:2014hca}       & NNPDF2.3LO~\cite{Ball:2012cx}     & \PYTHIAV{8.2}~\cite{Sjostrand:2014zea}    & A14     & NLO    \\
$\ttbar$           & \POWHEGBOX~\cite{Alioli:2010xd}               & NNPDF2.3LO~\cite{Ball:2012cx}     & \PYTHIAV{8.2}~\cite{Sjostrand:2014zea}    & A14     & NNLO+NNLL~\cite{Cacciari:2011hy,Czakon:2011xx} \\
Single top         & \POWHEGBOX~\cite{Alioli:2010xd}               & NNPDF2.3LO~\cite{Ball:2012cx}     & \PYTHIAV{8.2}~\cite{Sjostrand:2014zea}    & A14     & NNLO/NLO+NNLL~\cite{Campbell:2020fhf,Kidonakis:2011wy,Kidonakis:2010ux,Kidonakis:2010tc,Kidonakis:2021vob}      \\
Multiboson         & \SHERPA~2.2.1--2.2.2~\cite{Gleisberg:2008ta}  & NNPDF3.0NNLO~\cite{Ball:2014uwa}  & \SHERPA       & Default & NLO \\
\bottomrule
\end{tabular}
\vspace{3mm}
\label{tab:mcsamples}
\end{table}

\label{subsec:datasets:common}

Samples modeling the single production of~~$T$ quarks were generated
using the \MGMCatNLO program~\cite{Alwall:2014hca} at LO in quantum chromodynamics (QCD) for matrix elements (ME) with
the ``VLQ'' Universal FeynRules Output (UFO) model~\cite{UFO}
implementing the Buchkremer model introduced in \Refn{\cite{Buchkremer:2013bha}} and Eq.~(\ref{eqn:SimpLag}).
The generated events were interfaced with \PYTHIAV{8.2} using the NNPDF23LO PDF set and the A14 tune.
This model uses the 4-flavor scheme and all tree-level processes
are included. The VLQs are assumed to couple exclusively to SM quarks of the third generation and SM bosons.
In these analyses the pair production of VLQs is not considered.

Separate samples were produced for the \WTHt, \WTZt, \ZTHt and \ZTZt processes.
Samples were generated with a 200 \gev\ spacing for \WTHt and \WTZt processes, and
ME-based weights
were used to reweight the samples to intermediate mass points to create 100 \gev\ intervals
in the mass grid. Since the relative variations in the resonance lineshape of \ZTHt\ and \ZTZt\
samples are larger as a function of the $T$ quark mass, the reweighting requires additional mass points for
the precision desired, and samples for these processes were generated  with 100 \gev\ mass spacing.

For similar reasons, the universal coupling between the $T$ quark and the gauge boson, $\kappa$,
was set to a value of 1.0 for the \WTHt\ and \WTZt\ samples, and reweighted to other values
of $\kappa$ down to a value of 0.1, while this reweighting for the \ZTHt\ and \ZTZt\ samples is
only valid within much narrower ranges of $\kappa$. Thus, for the latter two processes,
samples were produced at $\kappa$ values of 0.2, 0.4, 0.7 and 1.0, and used to reweight to
neighboring coupling values.

Benchmark samples with specific values of the relative coupling parameters $\vec{\xi}$
(where $\vec{\xi} = (\xi_{W}, \xi_{Z}, \xi_{H}) = (0.50, 0.25, 0.25)$ for
\SU2 singlets in the high-mass asymptotic limit)\footnote{The parameterization of VLQ Lagrangian in terms of the $\kappa$ and $\vec{\xi}$ parameters was
introduced in Ref.~\cite{Roy:2020fqf} and is used for the interpretation presented. The conversion to the coupling convention in Eq.~(\ref{eqn:SimpLag}) is obtained through
a one-to-one mapping of the tree-level couplings in the Lagrangian.
The $\xi_{W}$, $\xi_{Z}$, $\xi_{H}$ parameters satisfy the constraint $\xi_{W} + \xi_{Z} + \xi_{H} = 1$
and represent the branching ratios of $T$ decaying into $Wb$, $Zt$, and $Ht$ in
the narrow-width approximation.}
were constructed by combining the samples for individual production
and decay modes by their appropriate relative proportions.
The samples were normalized by multiplying the LO cross-section times branching ratio for given assumed couplings
by a correction factor to account for finite width effects~\cite{Roy:2020fqf, Roy:2022wjj}, and by next-to-leading-order (NLO) $k$-factors computed in
the narrow-width approximation~\cite{Cacciapaglia:2018qep}. A change in the dynamic scale in \MADGRAPH at the
\VLTa width-over-mass threshold $\RW = 0.1$, leads to a discontinuity in the computed cross-section~\cite{Roy:2022wjj}. As a result, two different parameterizations of the cross-section are
available: $\sigma_{\mathrm{low}}(\RW)$ for the $\RW < 0.1$ regime, and $\sigma_{\mathrm{high}}(\RW)$
for $\RW > 0.1$.
An averaging procedure is used to obtain a smooth cross-section $\sigma(\Gamma/M)$ across the mass and
coupling grid:\\
\[
\sigma(\RW)=
\begin{cases}
\sigma_{\mathrm{low}}(\RW) + \frac{1}{2}[\sigma_{\mathrm{high}}(0.1) - \sigma_{\mathrm{low}}(0.1)],& \text{if } \RW < 0.1\\
\sigma_{\mathrm{high}}(\RW) - \frac{1}{2}[\sigma_{\mathrm{high}}(0.1) - \sigma_{\mathrm{low}}(0.1)],&  \text{if } \RW \geq 0.1.
\end{cases}
\]

An additional uncertainty of $\frac{1}{2}[\sigma_{\mathrm{high}}(0.1) - \sigma_{\mathrm{low}}(0.1)]$ is
assigned on the cross-section at every point to account for this choice.

%

%
%
%
%
%

%
%
%
%
%

%
%
%
%

%

% End of text imported from the .//Sections/DataAndSimulation.tex input file

% The next lines are included from the .//Sections/Analyses.tex input file
%
%

\section{Description of input analyses}

The analyses search for the production of charged $+2e/3$
vector-like $T$ quarks that decay into a $H$ boson and a top quark ($T \rightarrow Ht$) or a $Z$ boson and a top quark ($T \rightarrow Zt$).
The targeted vector-like $T$ production topology is displayed in Figure~\ref{fig:singleT}. Four production and decay modes are possible:
$ Wb \rightarrow T \rightarrow Ht$   %
, $ Wb \rightarrow T \rightarrow Zt$ %
, $ Zt \rightarrow T \rightarrow Ht$ %
and $ Zt \rightarrow T \rightarrow Zt$.   %
The final state is
characterized by the presence of multiple ($b$-tagged) jets due to the SM quarks and bosons produced, along with the recoiling
initial-state quark which typically manifests as a forward jet. Due to the difference between the masses of the top and bottom
quarks, $b$-associated (or $W$-mediated) \VLTa production is kinematically favored over $t$-associated (or $Z$-mediated)
production. However, in certain gauge representations, such as with a $(T,B)$
or an $(X,T)$ \SU2 doublet, the coupling to $W$ bosons vanishes, and the $t$-associated mode is the only allowed production channel.

\begin{table}[tb]
\begin{center}\vspace{0.3cm}
\caption{Summary of the target signal and decay channels for the three analyses included in the combination and the discriminating variables used.}
\begin{tabular}{cccc}
\toprule{Analysis} & {Target signal} & {Decay channels} & {Discriminants}\\
\midrule
\MONO   & $Wb/Zt   \rightarrow T \rightarrow Zt$     & $ Zt   \rightarrow \nu \nu bqq$ (0$\ell$) & BDT score\\
\HTZT     & $Wb/Zt   \rightarrow T \rightarrow Ht/Zt$  & $ Ht/Zt \rightarrow bbb\ell\nu/qqb\ell\nu$ (1$\ell$) & $m_{\mathrm{eff}}$ \\
\OSML      & $Wb/Zt   \rightarrow T \rightarrow Zt$     & $ Zt   \rightarrow \ell\ell b\ell\nu $ (3$\ell$), $Zt \rightarrow \ell\ell bqq$ (2$\ell$) & $Z$ boson \pT \\
\bottomrule
\end{tabular}
\label{tab:shortsummary}
\vspace{-0.5cm}
\end{center}
\end{table}

Full descriptions of the event selection for the different analyses
are available in Refs.~\cite{HtZtPaper2023,OSMLPaper2024,EXOT-2022-40}. A summary of the
target signals, decay channels and discriminants of each analysis
is given in Table~\ref{tab:shortsummary}. Common features
include the requirement of significant energy (or missing transverse momentum) observed in the detector
as leptons (electrons or muons) and jets.
Jets are reconstructed by applying the anti-$k_t$ algorithm~\cite{Cacciari:2008gp, Fastjet}
to topological calorimeter-energy clusters and charged-particle tracks processed with
a particle-flow algorithm~\cite{PERF-2015-09} with fixed radius parameters of $R=0.4$ (small-radius jets) and
$R=1.0$ (large-radius jets), and with a variable-radius parameter~\cite{Krohn:2009zg} optimized for top-quark tagging.
Small-radius jets may be central ($ | \eta | < 2.5$) or forward
($ 2.5 < | \eta | < 4.5$). Central jets may be $b$-tagged, while forward jets are a distinctive
signature of singly produced $T$ quarks, where the jet scatters off of a vector boson from one of the
incoming partons. By design, these analyses have orthogonal selections. This is achieved by
requiring exactly zero leptons in the \MONO analysis, exactly one lepton in the \HTZT analysis,
and two or more leptons in the \OSML analysis. As a result, there are no overlapping events in the signal regions.

The \MONO analysis~\cite{EXOT-2022-40} focuses on events with a reconstructed boosted
hadronically-decaying top quark
produced in association with large missing transverse momentum and a forward jet. The output score of a Boosted
Decision Tree (BDT) algorithm is used in the definition of the signal, control and validation regions, and is the discriminating variable in the signal regions.

The \HTZT search~\cite{HtZtPaper2023} analyzes data with final states containing
a single lepton with multiple jets including $b$-jets and a forward jet. The presence of boosted heavy resonances
in the event is exploited to discriminate the signal from the SM
background. Due to the strong signal discrimination power, the ``effective mass,''
$m_{\mathrm{eff}}$, defined as the scalar sum of the transverse momentum (\pT) of all central small-radius
jets and leptons in the event and the missing transverse energy, $\met$, is chosen as the
observable. The analysis is split into five signal regions based on jet multiplicity
to further enhance the sensitivity.

The final state of the \OSML analysis~\cite{OSMLPaper2024} is characterized
by the presence of a pair of electrons or muons with opposite-sign charges that forms a reconstructed $Z$ boson
candidate, and by the presence of $b$-tagged jets and a forward jet. Events with exactly
two or at least three leptons are categorized into two independently optimized analysis channels. The
transverse momentum of the $Z$ boson is chosen as the observable due to its discriminating power.

% End of text imported from the .//Sections/Analyses.tex input file

%
%

% The next lines are included from the .//Sections/StatisticalAnalysis.tex input file
\section{Statistical analysis and systematic uncertainties}
\label{section:stat_analysis}

For each benchmark scenario considered in this search, the distributions of the
discriminating variables of each analysis across their respective search regions
are jointly analyzed to test for the presence of signal. A combined likelihood
function is constructed as a product of Poisson probability terms over all
bins considered in each analysis. Systematic uncertainties are
implemented as nuisance parameters with constraints described by Gaussian distributions.

The three analyses take different approaches to model their respective backgrounds with different
associated systematic uncertainties. Although they
explore significantly different phase spaces, $\ttbar$ (and~$t\bar{t}V$), $Z+$jets and diboson production
dominate the background in the signal
regions. Each analysis applies a different background treatment, and as the specific background contributions
vary, the normalization factors are also determined independently. As a result, uncertainties in the theoretical modeling
and normalization of the background processes in each of the analyses are
assumed to be uncorrelated as each analysis probes a significantly different region of phase space.

Detector-related uncertainties are applied, including those pertaining to the lepton measurement and identification,
missing transverse momentum measurement, and jet-related quantities. These uncertainties are applied to both signal
and background samples and are correlated where similar selection criteria
are used (see Table~\ref{tab:statana_syst_table}). Furthermore, each of the analyses
employs $b$-tagging to enhance the signal separation. In the \MONO and \OSML analyses, the $b$-tagging implementation is identical and the associated systematic
uncertainties are correlated. The \HTZT analysis uses track-based jets instead of calorimeter-based jets
and therefore the
systematic uncertainties are left uncorrelated between this channel and the other two.
In practice the choice of decorrelating this channel is observed to have
no impact on the result.

Similar to the $b$-tagging uncertainties, jet-related systematic uncertainties are
considered to be correlated only between analyses that use the same jet definition. There
are three exceptions. First, the jet energy resolution (JER) uncertainty is decorrelated between
the \MONO analysis and the others, as the former uses a more detailed scheme for this uncertainty, which was not necessary for the other analyses. Decorrelating results in a larger, more conservative uncertainty but has no impact on the combination result.
Although this is a conservative choice, in practice it is observed to have no impact on the result
due to the insensitivity of the analyses to these jet uncertainties. Second, the jet
energy scale (JES) uncertainties for small-radius jets and large-radius jets are
partially correlated. This is implemented via a set of nuisance parameters, a subset of
which are common to both types of jets, while other nuisance parameters in the set
apply only to one of the two types of jets. Third, the jet mass resolution (JMR) uncertainties
are uncorrelated between the \HTZT and \MONO analyses (and are not present in
the \OSML analysis). The \MONO analysis uses JMR uncertainties related to large-radius jets,
while the \HTZT analysis uses JMR uncertainties related to small-radius jets.

Detector uncertainties affecting the signal acceptance are
considered when setting exclusion limits, but uncertainties affecting the production cross-section or signal shape are not.
However, the uncertainty in the theoretical signal cross-section prediction (Section~\ref{sec:datatrig}) is shown
in Figures~\ref{fig:lims-singlet} and \ref{fig:lims-doublet}.

Uncertainties due to the limited size of the simulated samples are taken into account
with dedicated parameters in the fit that are independent across the bins. These
parameters are included with Poisson constraints.
Finally, a set of unconstrained parameters are included to control the normalization
of certain background processes in each analysis. Table~\ref{tab:statana_syst_table} summarizes the
systematic uncertainty model used in the fit.

The likelihood function depends on the signal-strength parameter $\mu$, which
multiplies the predicted production cross-section for signal, and $\theta$, the set
of nuisance parameters that encode the effect of systematic uncertainties in the
signal and background expectations. Therefore, the expected total number of events
in a given bin depends on $\mu$ and $\theta$.

The test statistic $q_{\mu}$ is defined as the profile likelihood ratio:
$q_\mu = -2\ln({\cal L}(\mu,\hat{\hat{\theta}}_\mu)/{\cal L}(\hat{\mu},\hat{\theta}))$. Here,
$\hat{\mu}$ and $\hat{\theta}$ are the values of the parameters $\mu$ and $\theta$
that simultaneously maximize the likelihood function ${\cal L}(\mu,\theta)$ (subject
to the constraint $0\leq \hat{\mu} \leq \mu$), whereas $\hat{\hat{\theta}}_\mu$ are the
values of the nuisance parameters that maximize the likelihood function for a given value
of $\mu$. The statistic used for the discovery test, to compute the compatibility of the observed data
with the background-only hypothesis, is obtained by setting $\mu=0$ in
the profile likelihood ratio and leaving $\hat{\mu}$ unconstrained:
$q_0 = -2\ln({\cal L}(0,\hat{\hat{\theta}}_0)/{\cal L}(\hat{\mu},\hat{\theta}))$.
The $p$-value of the discovery test is given by the integral of the probability
distribution of $q_{0}$ above the observed $q_{0}$ value when assuming the
background-only hypothesis, and it is computed using the asymptotic approximation detailed in
Ref.~\cite{Cowan:2010js}. For each signal scenario considered,
the upper limit on the signal production cross-section is computed using $q_\mu$ in
the CL$_{\textrm{s}}$ method~\cite{Junk:1999kv,0954-3899-28-10-313}, also in the asymptotic
approximation. For a given signal scenario, values of the production cross-section
(parameterized by $\mu$) yielding CL$_{\textrm{s}} < 0.05$ are excluded at $\geq 95\%$ CL.
The combination is performed with RooStats~\cite{moneta2011roostats} with statistical models implemented using RooFit~\cite{verkerke2003roofit}
and HistFactory~\cite{CERN-OPEN-2012-016}.

\begin{table}[tb]
\caption{Summary of systematic uncertainties and normalization factors included
in the combined likelihood fit. A \checkmark indicates the uncertainty is
included in a specific channel. The normalizations of the largest backgrounds in the \HTZT analysis are constrained with nuisance parameters subject to Gaussian constraints. %
}
\begin{tabular}{p{0.3\linewidth} |p{0.11\linewidth} |p{0.11\linewidth} |p{0.11\linewidth} |p{0.2\linewidth} }
\toprule

Category & \MONO & \HTZT & \OSML & Correlating \\
\hline
\multicolumn{5}{c}{Lepton and \met uncertainties}\\
\hline
Electron uncertainties &  & $\checkmark$ & $\checkmark$ & All \\
Muon uncertainties &  & $\checkmark$ & $\checkmark$ & All \\
\hline
\met uncertainties & $\checkmark$ & $\checkmark$ & $\checkmark$ & All \\
\hline
\multicolumn{5}{c}{Jet uncertainties}\\
\hline
JES uncertainties & $\checkmark$ & $\checkmark$ & $\checkmark$ & All \\
JER uncertainties & $\checkmark$ & $\checkmark$ & $\checkmark$ & \HTZT and \OSML\\
JMS uncertainties &  & $\checkmark$ &  & None \\
JMR uncertainties & $\checkmark$ & $\checkmark$ & & None \\\hline
\multicolumn{5}{c}{Tagging uncertainties}\\
\hline
Flavor-tagging uncertainties & $\checkmark$ & $\checkmark$ & $\checkmark$ & \MONO and \OSML\\
Top-tagging uncertainties & $\checkmark$ & & & None \\
$W$/$Z$-tagging uncertainties & $\checkmark$ & & & None \\
\hline
Background modeling uncertainties (constrained) & $\checkmark$ & $\checkmark$ & $\checkmark$ & None \\
\hline
\multicolumn{5}{c}{Background normalization factors (unconstrained)}\\
\hline
$\ttbar$ normalization & $\checkmark$ & & & None \\
$V$+jets normalization & $\checkmark$ & & & None \\
$Z$+light-jets normalization & & & $\checkmark$ & None \\
$Z$+heavy-flavor normalization & & & $\checkmark$ & None \\
$t\bar{t}V$ normalization & & & $\checkmark$ & None \\
$VV$ normalization & & & $\checkmark$ & None \\
\bottomrule

\end{tabular}
\label{tab:statana_syst_table}

\end{table}

% End of text imported from the .//Sections/StatisticalAnalysis.tex input file

% The next lines are included from the .//Sections/Results.tex input file
\section{Results}
\label{sec:results}

No significant excess above the SM expectations is observed in any of the analysis channels
considered for the combination. A signal search is performed using the combined analyses and
no significant excess is found. Scanning across the $\MT$ and $\kappa$ parameters, the most significant local $p_0$-value of 0.14 (0.10) for the \SU2 singlet (doublet) interpretation is found for the signal point $\MT= 2.1$~TeV and $\kappa = 0.1$.
Good agreement with SM predictions is also found in the combined background-only fit to data.
Therefore, upper limits on the cross-section are set at $95\%$~CL as a function of $\MT$ and $\kappa$. The limits are calculated for the sum of the
production cross-sections times branching ratio of the four production and decay modes considered and denoted as $\sigma(pp \to T \to Zt/Ht)$.
In contrast, the \OSML and \MONO analyses were not designed to be sensitive to the $T \rightarrow Ht$ decay mode, and thus the corresponding publications reported upper limits on a subset of the processes ($T \to Zt$). %

By comparing the obtained cross-section limits with the theoretical cross-section, limits are derived on $\MT$ and $\kappa$. As the signal efficiencies for the considered signal
modes are generally different, limits are independently determined for combinations of $\MT$,
$\kappa$, and branching ratios. Motivated by the Goldstone equivalence theorem~\cite{Chanowitz:1985hj},
which states that the branching ratios of $T$ quark decaying into $Zt$ and $Ht$ become
similar in the large-$\MT$ limit under the narrow-width approximation, $\xi_Z = \xi_H$ is required.

The expected and observed limits on the $T \rightarrow Ht/Zt$ cross-section from
the combination are presented for several illustrative benchmark points
for the \SU2 singlet (with a branching ratio $T\rightarrow Wb$ of 0.5) and doublet (with a branching ratio $T\rightarrow Wb$ of 0.0) representations. \Fig{\ref{fig:lims-singlet}} (\Fig{\ref{fig:lims-doublet}}) shows the
limits on the total cross-section of the sum of the production and decay processes
for the \SU2 singlet (doublet) representation assuming $\kappa = 0.3, 0.5$.

Following Ref.~\cite{Deandrea:2021vje}, this interpretation is only used
where $\RW$ is smaller than 50\%, the region in which the correction factors for the
finite-width approximation and the non-resonant contributions~\cite{Roy:2020fqf, Roy:2022wjj}
are also valid. In \Fig{\ref{fig:lims-K050-singlet}} and
\Fig{\ref{fig:Lims-K050-doublet}}
a vertical dashed line is drawn to indicate the lower edge of the region with $\RW > 50\% $.
Such lower edge extends to lower mass values
as the $\kappa$ value increases. As a result, the highest mass for
which the limit is shown decreases as the value of $\kappa$ increases.

\begin{figure}[tb]
\centering
\subfloat[]{
\includegraphics[width=0.47\textwidth]{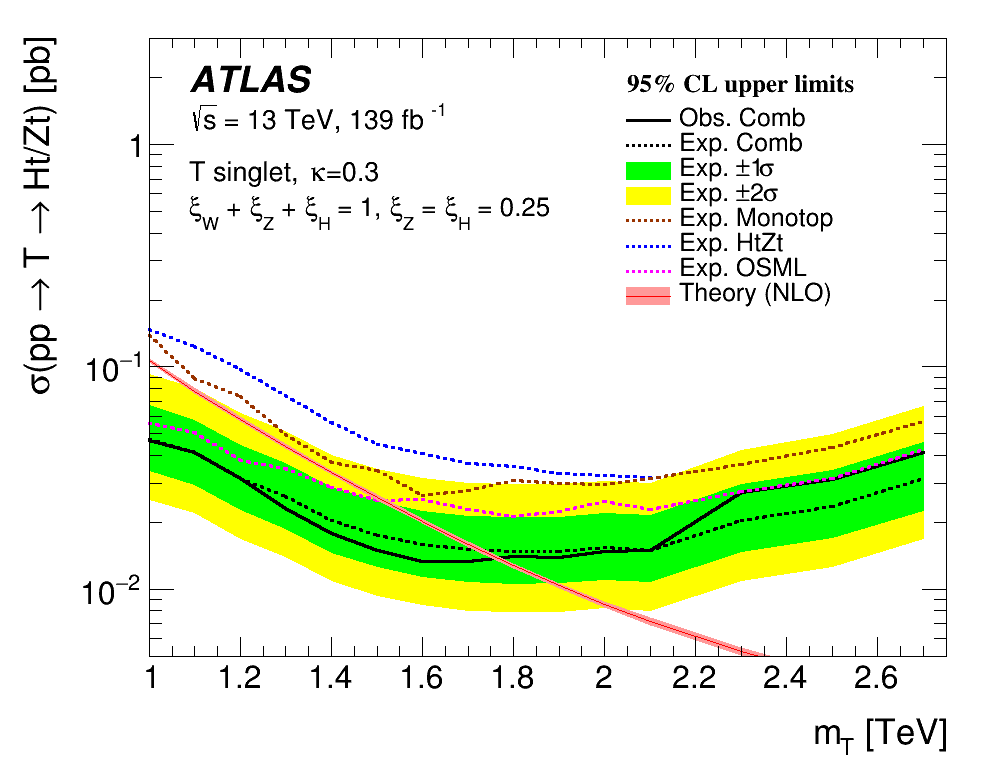}
\label{fig:lims-K030-singlet}
}
\subfloat[]{
\includegraphics[width=0.47\textwidth]{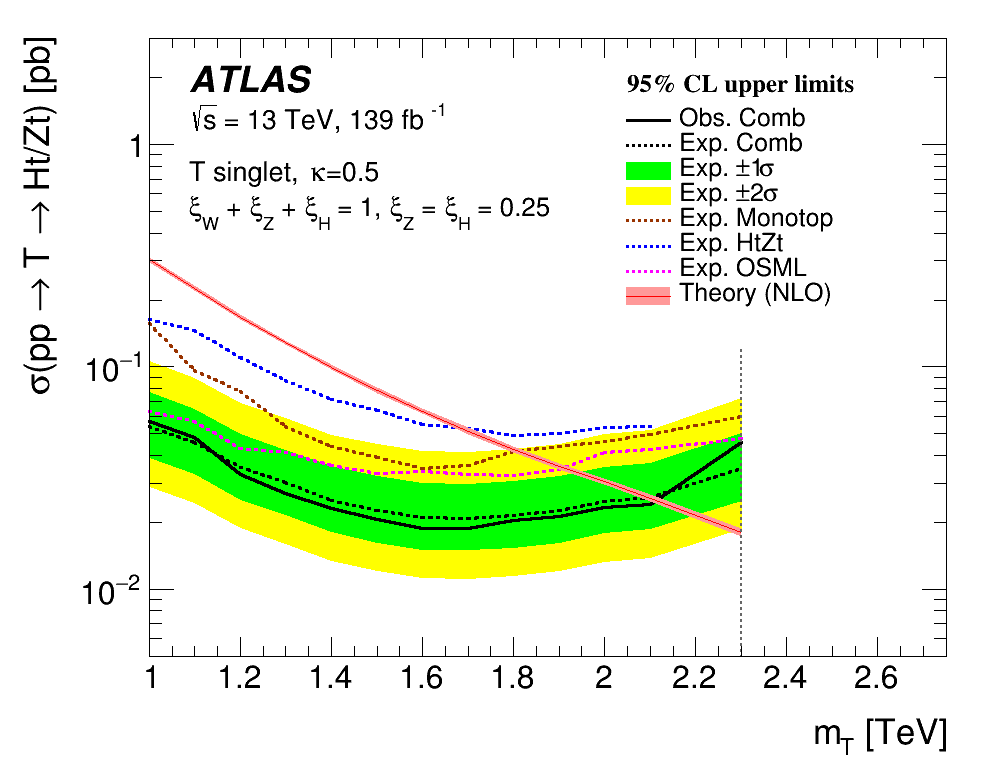}
\label{fig:lims-K050-singlet}
}

\caption[]{
Observed (solid line) and expected (dashed line) 95\% CL upper limits on the total cross-section $\sigma(pp \to T \to Ht/Zt)$ as a function of \VLTa mass in the \SU2 singlet representation assuming \protect\subref{fig:lims-K030-singlet} $\kappa=0.3$ and \protect\subref{fig:lims-K050-singlet} $\kappa=0.5$.  The surrounding shaded bands correspond to  1 and  2 standard deviations around the expected limit. A vertical dashed line is drawn to indicate the lower edge of the region with $\RW  > 50\% $ for which the theoretical calculations are no longer
valid. The expected limits for the individual analyses are shown. The \HTZT analysis is only included in the limit calculation for $\MT < 2.1$~TeV. The red line shows the NLO theoretical cross-section prediction, with the surrounding shaded band representing the corresponding uncertainty.
}
\label{fig:lims-singlet}
\end{figure}

\begin{figure}[tb]
\centering
\subfloat[]{
\includegraphics[width=0.47\textwidth]{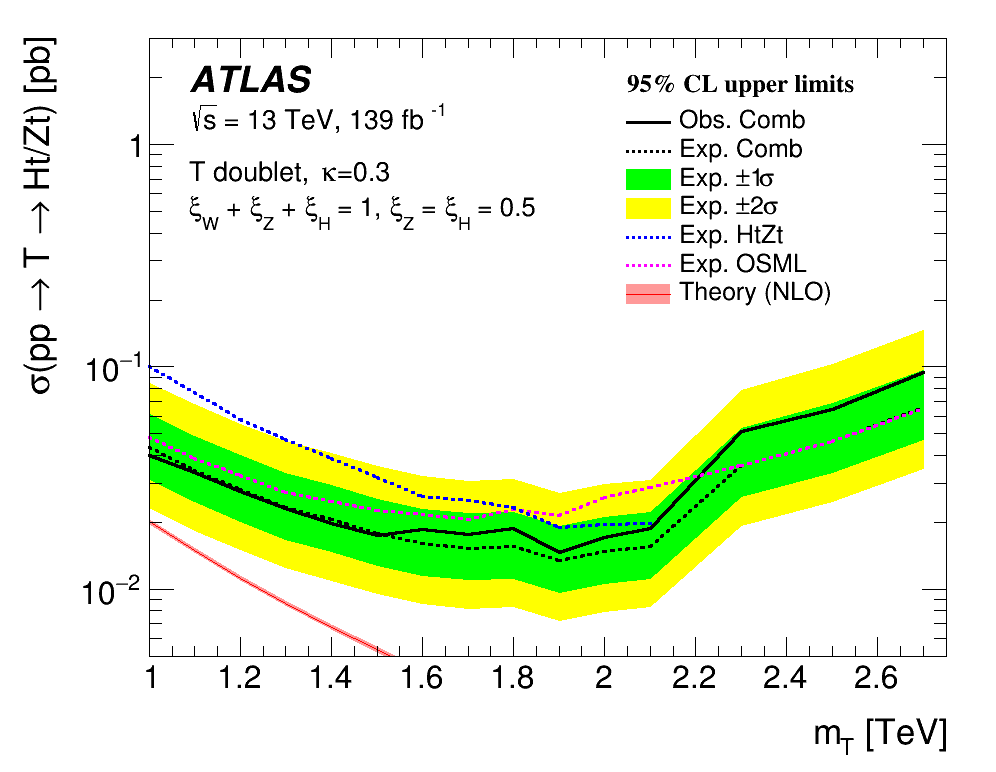}
\label{fig:Lims-K030-doublet}
}
\subfloat[]{
\includegraphics[width=0.47\textwidth]{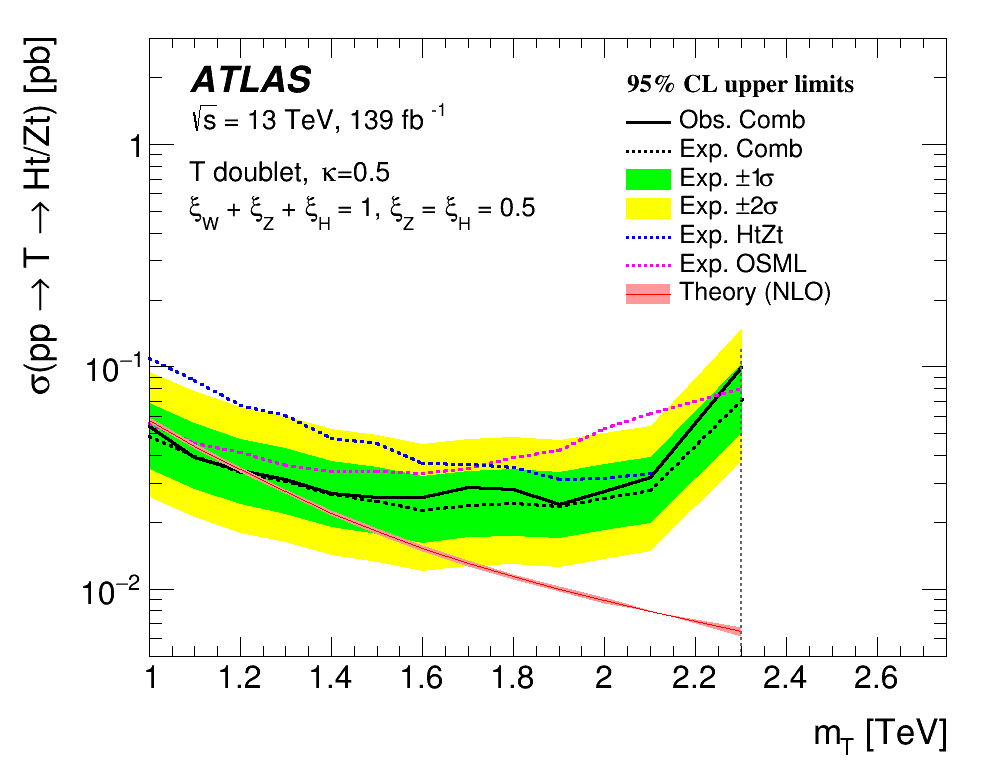}
\label{fig:Lims-K050-doublet}
}
\caption[]{Observed (solid line) and expected (dashed line) 95\% CL upper limits on the total cross-section $\sigma(pp \to T \to Ht/Zt)$ as a function of \VLTa mass in the \SU2 doublet representation assuming \protect\subref{fig:Lims-K030-doublet} $\kappa=0.3$ and \protect\subref{fig:Lims-K050-doublet} $\kappa=0.5$.  The surrounding shaded bands correspond to  1 and 2 standard deviations around the expected limit. A vertical dashed line is drawn to indicate the lower edge of the region with $\RW  > 50\% $ for which the theoretical calculations are no longer
valid. The expected limits for \HTZT and \OSML analyses are shown;
the \MONO analysis is not shown due to substantially less sensitivity in this scenario.
However, the \MONO analysis is included in the combined limit calculation. The \HTZT analysis is only included in the limit calculation for $\MT < 2.1$~TeV. The red line shows the NLO theoretical cross-section prediction, with the surrounding shaded band representing the corresponding uncertainty.
}
\label{fig:lims-doublet}
\end{figure}

Results from the individual channels are overlaid on the combined results for comparison.
The statistical combination improves the limits over the individual results for all masses and couplings.
Existing limits from searches for VLQ pair production exclude \VLTa masses less than 1.5~\tev, with the
assumption of narrow-width resonances. These analyses do not consider pair production signals.
For the \SU2 doublet representation, the combination also includes the \MONO analysis, although its contribution is negligible. For this
reason, Figure~\ref{fig:lims-doublet} shows the limits from the \OSML and \HTZT analyses only.
In both Figures~\ref{fig:lims-doublet} and \ref{fig:lims-singlet} the \HTZT analysis is included in the combination for $\MT < 2.1$~TeV,
for consistency with the existing analysis.

The complementarity of the different analysis channels
is also evident in Figures~\ref{fig:lims-singlet} and~\ref{fig:lims-doublet}. For example, the \OSML analysis is most sensitive at low masses, while the
sensitivity of the \HTZT analysis can be seen to improve at higher masses. This is especially
true in the \SU2 doublet representation, as the \HTZT analysis includes signal regions that are specifically
designed to target $Z$-mediated production processes. The experimental sensitivity of the individual
channels and the combination depends on both the mass and width of the $T$ quark. In the \OSML analysis,
the mass dependence is clear, with the limit degrading above 2.1~TeV. This is due to the choice of binning of the
discriminant (the reconstructed $Z$ boson \pt), which was optimized to search for $T$ quarks with masses less than 2.0~TeV.
On the other hand, all three analyses use
discriminants that are relatively agnostic to the resonance width (e.g., the $m_{\mathrm{eff}}$ variable
for the \HTZT analysis, and the reconstructed $Z$ boson \pT in the \OSML analysis). Thus, the
excluded cross-section does not strongly depend on $\RW$ for the $\kappa$ values shown in Figures~\ref{fig:lims-singlet} and~\ref{fig:lims-doublet}.

The combination increases the sensitivity to a wider range of model parameters (generalized as the coupling $\kappa$) beyond the existing analyses.
The cross-section limits calculated for different choices
of coupling are interpreted as
exclusion limits computed as a function of $\MT$ and $\kappa$. \Fig{\ref{fig:Lims-singlet}} (\Fig{\ref{fig:Lims-doublet}}) shows
the observed and expected limits on the total cross-section for the \SU2 singlet (doublet) representation.
The limits are obtained for points on the $\MT$--$\kappa$ grid, spaced by 100~GeV in mass, and by 0.1 in $\kappa$.
All three channels are combined for masses up to 2.1~TeV.

\begin{figure}[tb]
\centering
\subfloat[]{
\includegraphics[width=0.47\textwidth]{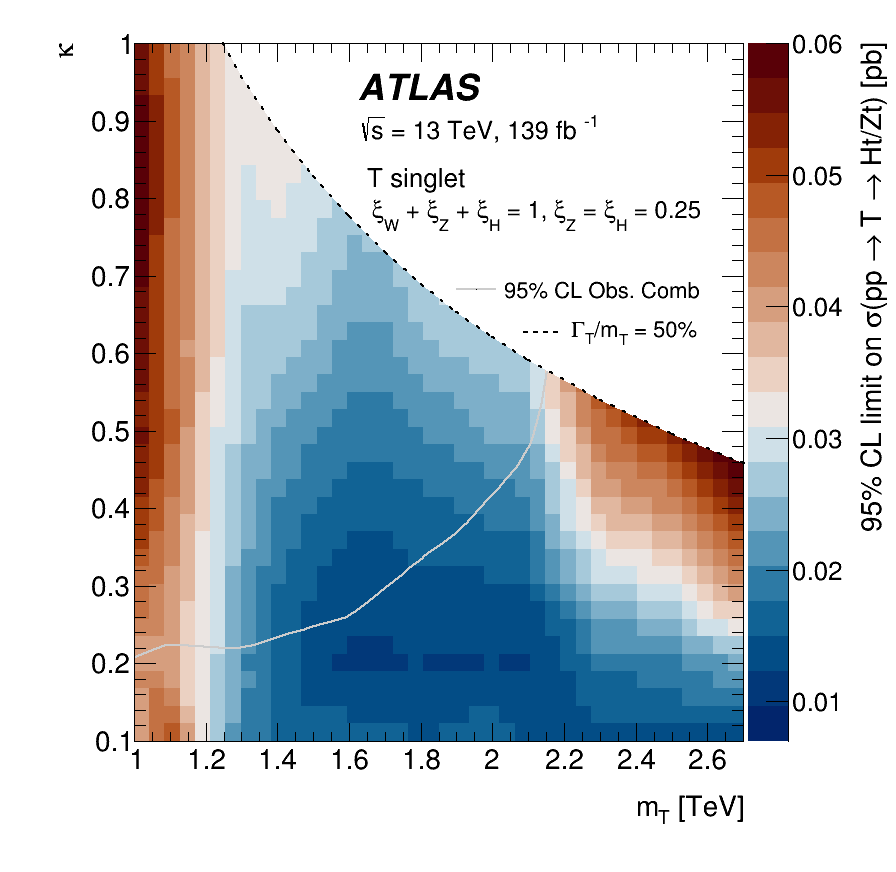}
\label{fig:XSLims-Obs-singlet}
}
\subfloat[]{
\includegraphics[width=0.47\textwidth]{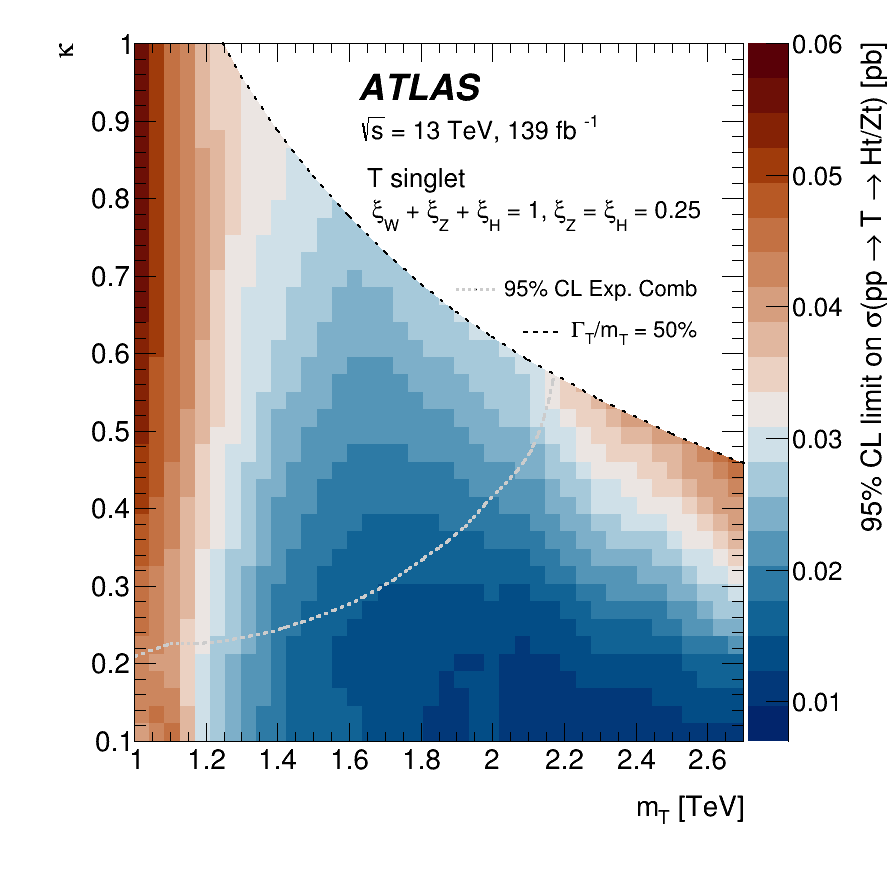}
\label{fig:XSLims-Exp-singlet}
} \\
\caption[]{
\protect\subref{fig:XSLims-Obs-singlet} Observed and \protect\subref{fig:XSLims-Exp-singlet} expected 95\% CL exclusion limits
on the total cross-section $\sigma(pp \to T \to Ht/Zt)$ as a function of the universal coupling constant $\kappa$ and the \VLTa mass in
the \SU2 singlet representation. All values of $\kappa$  above the white contour line are excluded at each mass point.
Limits are only presented in the regime $\RW < 50\%$, where the theory calculations are known to be valid.
}
\label{fig:Lims-singlet}
\end{figure}

\begin{figure}[tb]
\centering
\subfloat[]{
\includegraphics[width=0.47\textwidth]{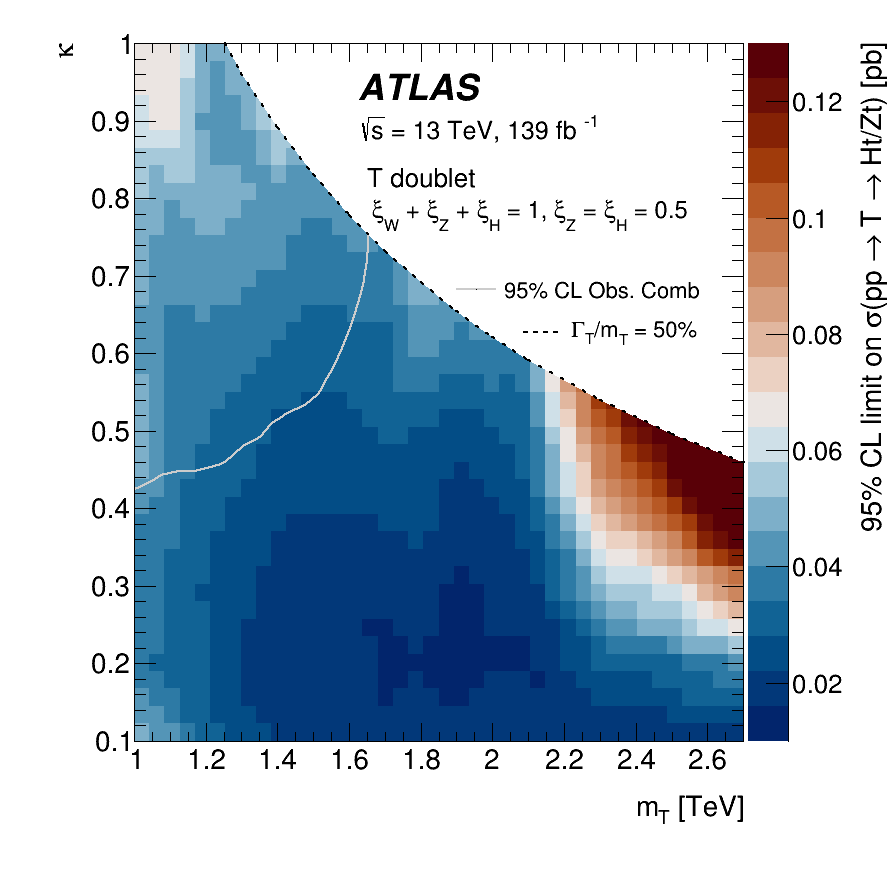}
\label{fig:XSLims-Obs-doublet}
}
\subfloat[]{
\includegraphics[width=0.47\textwidth]{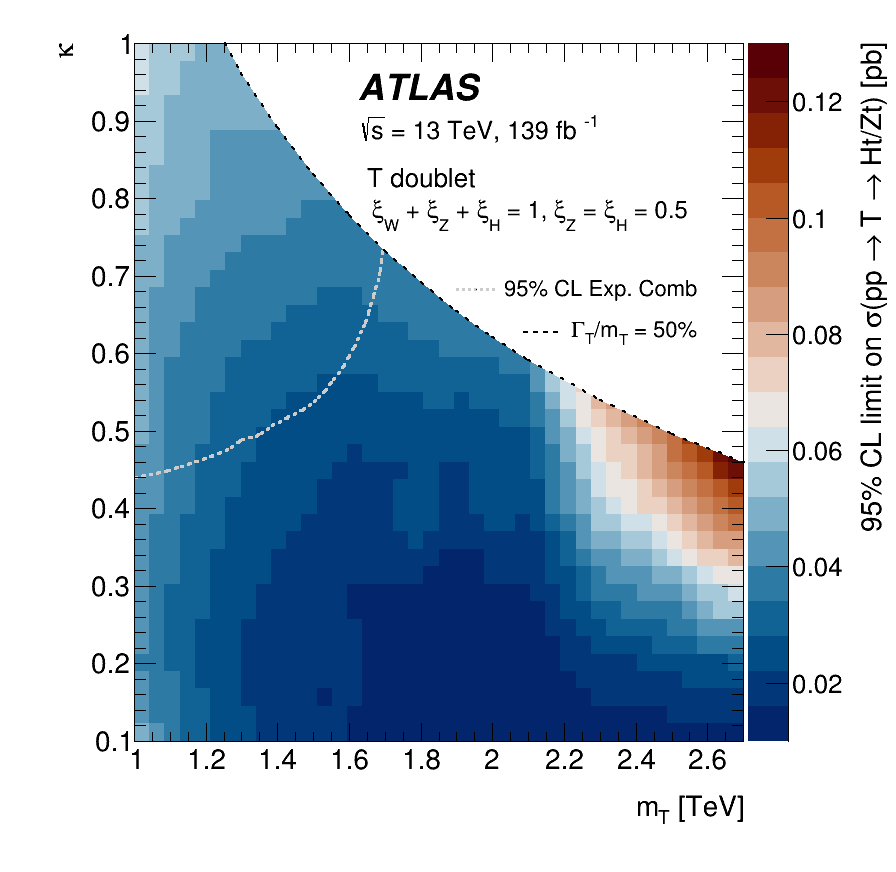}
\label{fig:XSLims-Exp-doublet}
} \\
\caption[]{
\protect\subref{fig:XSLims-Obs-doublet} Observed and \protect\subref{fig:XSLims-Exp-doublet} expected 95\% CL exclusion limits
on the total cross-section $\sigma(pp \to T \to Ht/Zt)$ as a function of the universal coupling constant $\kappa$ and the $T$ quark mass in
the \SU2 doublet representation. All values of $\kappa$  above the white contour line are excluded at each mass point.
Limits are only presented in the regime $\RW < 50\%$, where the theory calculations are known to be valid.
}
\label{fig:Lims-doublet}
\end{figure}

As a function of $\MT$ and $\kappa$, \Fig{\ref{fig:Lims-singlet}} and \Fig{\ref{fig:Lims-doublet}} show
the exclusion in a dashed line, where the region above and to the left (toward lower masses and larger couplings) is excluded.
Figure~\ref{fig:Lims-contour} shows the excluded regions in the $\MT$--$\kappa$ plane in more detail.
The limits are computed for a finite number of points in the $\MT$--$\kappa$ plane. As a result, the limits are interpolated using a piecewise function~\cite{2020SciPy-NMeth} between the
measured points to obtain a continuous shape for the exclusion  contours on the $\MT$--$\kappa$ plane. The
exclusion contours are shown both for the combination and for the individual
channels. Also shown in Figure~\ref{fig:Lims-contour} are the isolines corresponding to different values of $\RW$.
Limits are only shown for the parameter range where $\RW < 50\%$, where the theory calculations are known to be valid.

\begin{figure}[tb]
\centering
\subfloat[]{
\includegraphics[width=0.47\textwidth]{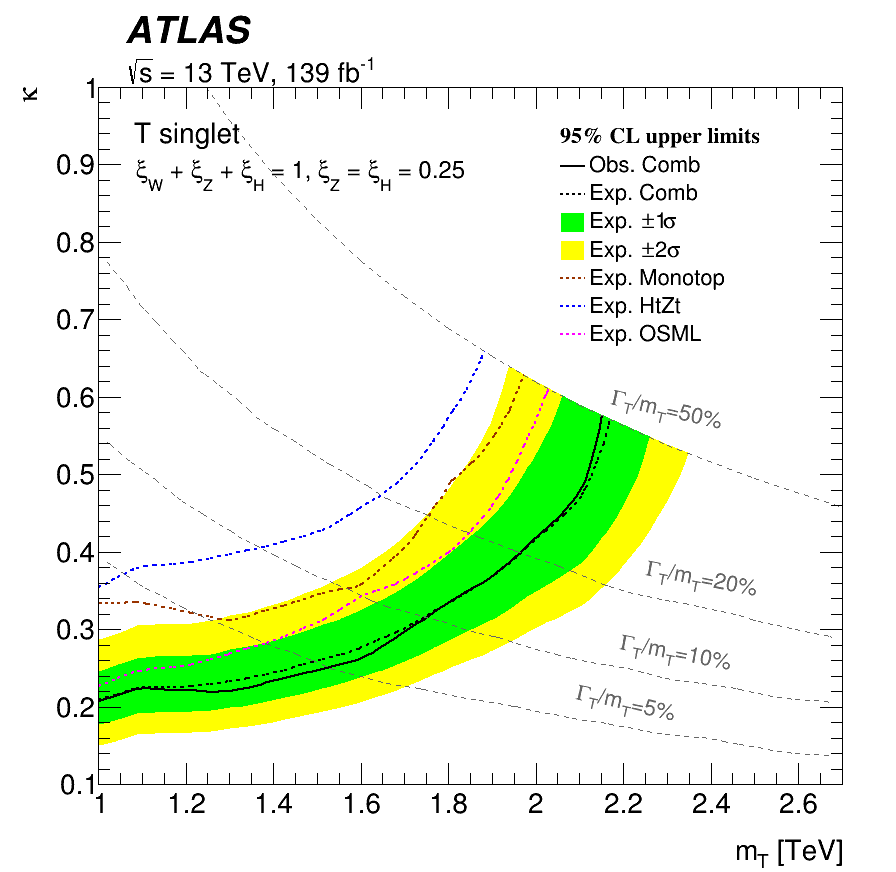}
\label{fig:MKLims-singlet}
}
\subfloat[]{
\includegraphics[width=0.47\textwidth]{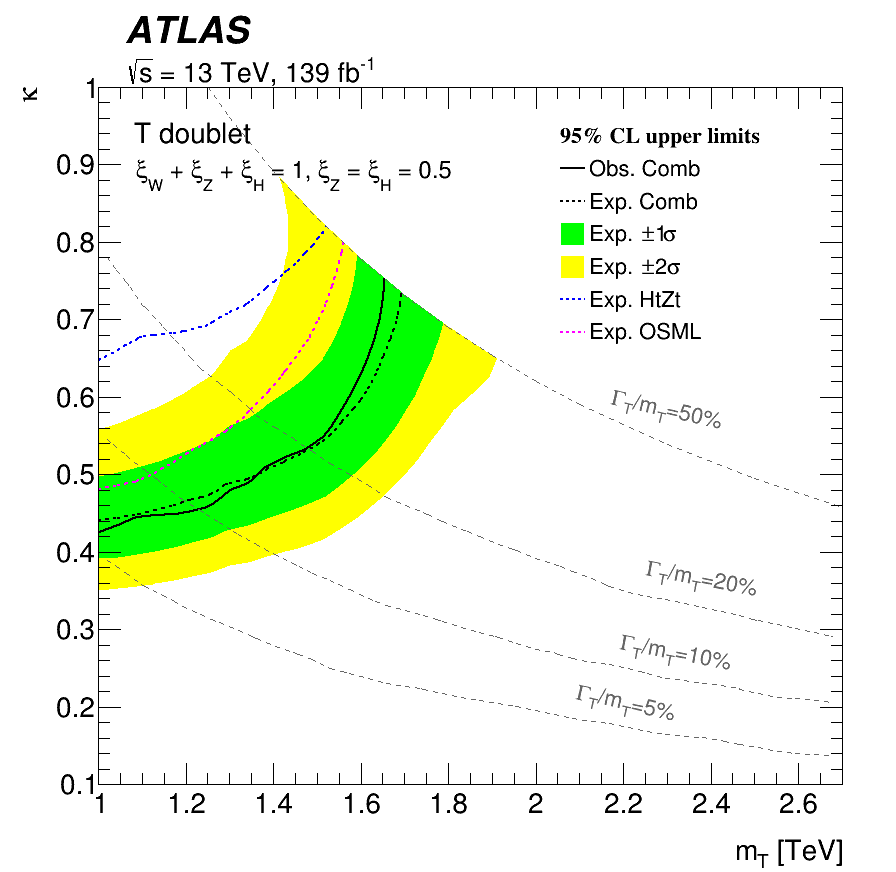}
\label{fig:MKLims-doublet}
}\\
\caption[]{
Observed (solid line) and expected (dashed line) 95\% CL exclusion limits on the universal coupling constant $\kappa$
as a function of the $T$ quark mass in the  \protect\subref{fig:MKLims-singlet} \SU2 singlet and \protect\subref{fig:MKLims-doublet} \SU2 doublet
representations for the combination. All values of $\kappa$ above the black contour lines are excluded at each mass point. The
shaded bands correspond to 1 and 2 standard deviations around the expected limit. Also shown are the expected
limits for the individual analyses. The gray dashed lines represent configurations of $(\MT, \kappa)$
resulting in equal values of the relative resonance width $\RW$. Limits are only presented in the regime $\RW < 50\%$,
where the theory calculations are known to be valid.
}
\label{fig:Lims-contour}
\end{figure}

The combination of the three searches significantly improves the exclusion limits from the individual channels.
In the \SU2 singlet representation, the coupling parameter $\kappa$ is constrained to be below 0.2 for the lower masses, and the mass is
excluded up to 2.1 TeV for $\kappa$ values near 0.6. The cross-section of a singly-produced $T$ quark is constrained to be
below $18$~fb for masses in the range of 1.4--2.2~TeV and $\kappa = 0.3$ for the \SU2 singlet representation. Similarly, for the \SU2 doublet representation the
coupling parameter $\kappa$ is constrained to be below 0.4 for the lower masses, and the mass is excluded up to 1.7~TeV
for $\kappa$ values near 0.7. The cross-section of a singly-produced $T$ quark is constrained to be below $20$~fb for
masses in the range of 1.4--2.1~TeV and $\kappa = 0.3$ in the \SU2 doublet representation. In both the \SU2 singlet and doublet representations, the observed cross-section
limits are slightly higher than the expected cross-section limits for $\kappa = 0.3$ and  $\kappa = 0.5$ for $\MT > 2.2$~TeV.
%

%

%
%

%
%
%

%

% The next lines are included from the .//Sections/Results_GammaBR.tex input file
%
%

The exclusion limits shown can be generalized for arbitrary values of
the $\xi_W$ parameter. Figure~\ref{fig:GammaBR} shows the observed and expected exclusions on the $T$ quark mass across the
plane spanned by $\xi_W$ and $\RW$. The relative width $\RW$ of the $T$ quark
is completely determined by the $\MT$ and $\kappa$ parameters. The results are shown under the
theoretically motivated assumption $\xi_Z$=$\xi_H$.

As expected from the previous results, the combination improves the exclusion limits from the individual channels.
The largest excluded mass is 2.1 TeV~for large $\RW$ and $\xi_W =0.5$. This is
equivalent to the \SU2 singlet representation, with a branching ratio to $Wb$ of 50\%. The lowest excluded masses
are observed for small $\RW$ (lower cross-section) near $\xi_W =0.0$ and $\xi_W =1.0$.
As $\xi_W$ approaches 0.0, the $T$ quark only decays into $Zt$ and $Ht$, while as $\xi_W$ approaches 1.0
there is no decay to either $Zt$ or $Ht$. The overview of benchmark mass limits
is shown in Table~\ref{tab:masslimits}.

\begin{figure}[tb]
\centering
\subfloat[]{
\includegraphics[width=0.50\textwidth]{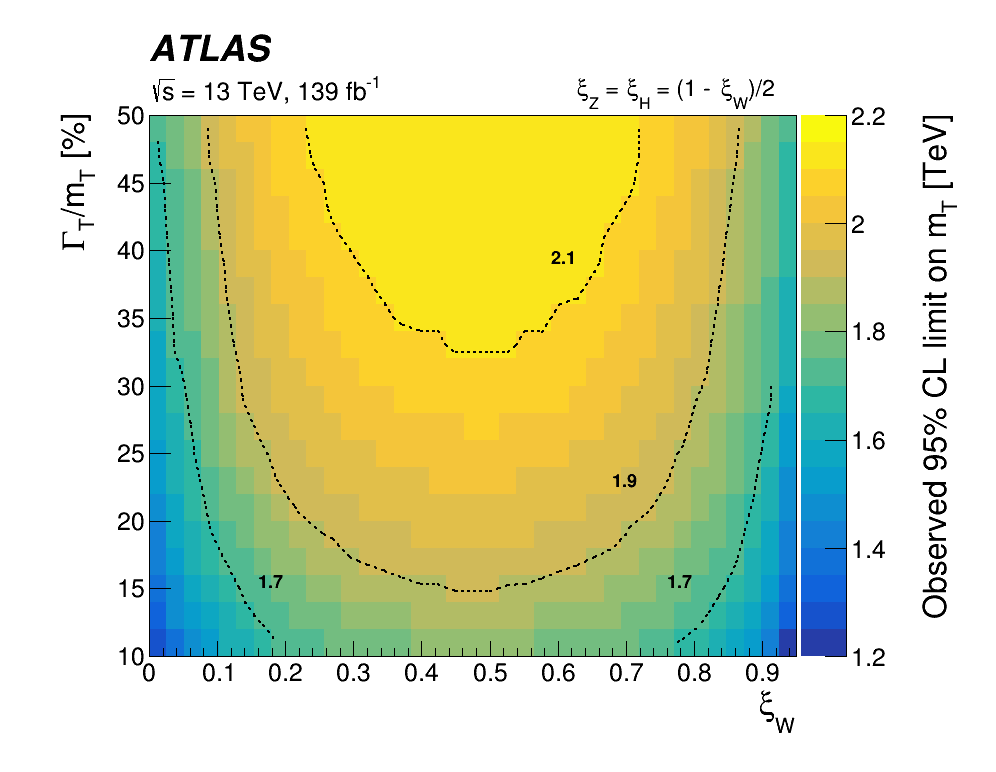}
\label{fig:Obs-GammaBR}
}
\subfloat[]{
\includegraphics[width=0.50\textwidth]{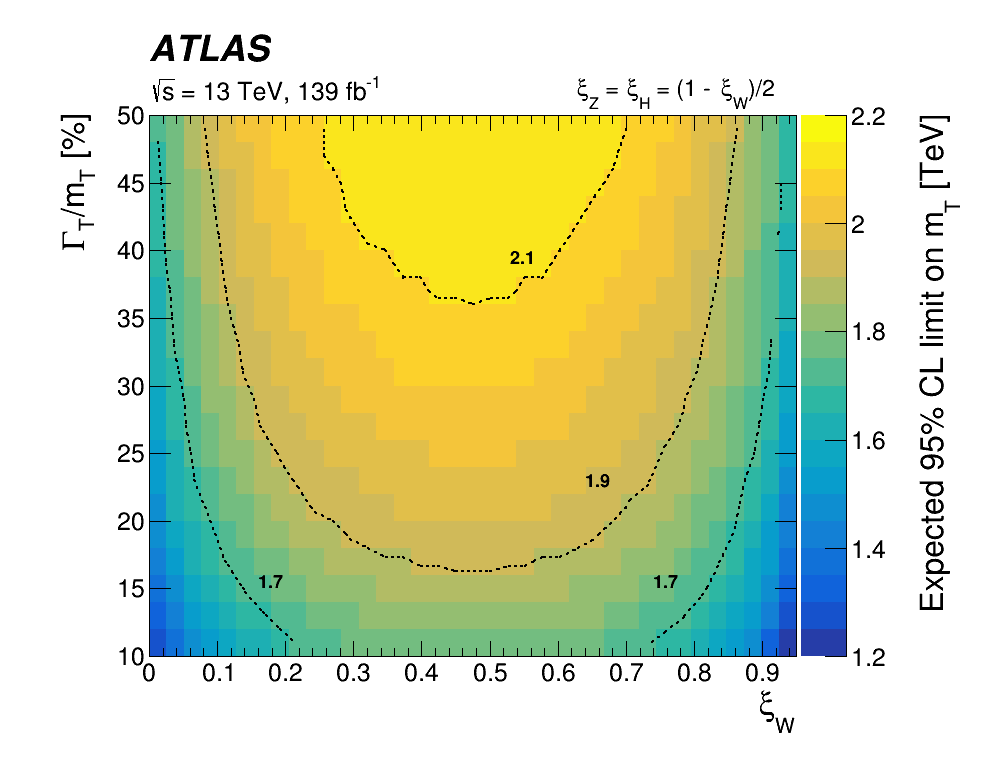}
\label{fig:Exp-GammaBR}
}
\caption[]{
\protect\subref{fig:Obs-GammaBR}  Observed and \protect\subref{fig:Exp-GammaBR} expected upper limits at 95\% CL
on the $T$-quark mass as a function of the relative resonance width ($\RW$) and the relative coupling parameter $\xi_W$,
for the assumption $\xi_Z=\xi_W$. The values $\xi_W=0.5$ and $\xi_W=0.0$ correspond
to the \SU2 singlet and \SU2 doublet representations, respectively.
The dashed contour lines denote exclusion limits of equal mass in units of TeV.
}
\label{fig:GammaBR}

\end{figure}

\begin{table}[tb]
\centering
\small  %
\caption{Summary of mass limits for \SU2 singlet and doublet representations with varying $\RW$ ratios. Both observed (Obs.) and expected (Exp.) limits are presented.}
\begin{tabularx}{0.85\textwidth}{@{\hspace{0.65cm}}l>{\centering\arraybackslash}X>{\centering\arraybackslash}Xc@{\hspace{0.65cm}}}
\toprule
Representation          & $\RW$ [\%]  & Obs./Exp. mass limit [TeV]   \\
\midrule
\SU2 singlet ($\xi_W = 0.5$) & 20                              & 2.0 / 2.0  \\
\SU2 singlet ($\xi_W = 0.5$) & 50                              & 2.1 / 2.1  \\
\SU2 doublet ($\xi_W = 0.0$) & 20                              & 1.4 / 1.4  \\
\SU2 doublet ($\xi_W = 0.0$) & 50                              & 1.6 / 1.7  \\
\bottomrule
\end{tabularx}
\label{tab:masslimits}
\end{table}

% End of text imported from the .//Sections/Results_GammaBR.tex input file

% End of text imported from the .//Sections/Results.tex input file
 %
\FloatBarrier

% The next lines are included from the .//Sections/Conclusions.tex input file
\section{Conclusions}

The first combination of results from the ATLAS Collaboration of searches for the single production of vector-like top quarks ($T$) decaying into $Ht$ or $Zt$ in hadronic and semileptonic final states has been presented.
Data from $pp$ collisions at $\sqrt{s}=13~\tev$ collected with the ATLAS detector at the LHC during 2015-2018 were used.
The dataset corresponds to an integrated luminosity of $139~\ifb$.
The observed data are consistent with the background prediction of the SM. Without a significant excess above the SM expectation, upper limits at 95\% CL are provided for a range of masses and couplings of the vector-like top quark, based on benchmark models and generalized representations. Additionally, the exclusion limits as a function of mass~($\MT$), coupling~($\kappa$), relative width~($\RW$) and branching ratio are provided.

The 95\% CL limits on the cross-section for \VLTa production in the \SU2 singlet representation are less than 47~fb (57~fb) for all $\MT$ between 1.0 and 2.7 (2.3)~TeV for coupling $\kappa = 0.3$ (0.5), with a limit of 13~fb for $\kappa = 0.3$ and $\MT = 1.6$~TeV. In the \SU2 doublet representation, the 95\% CL limits on the cross-section are less than 95~fb (100~pb) over the same mass region and coupling values, with a limit of 15~fb for $\kappa = 0.3$ and $\MT = 1.9$~TeV. The combined results improves the cross-section limits over the individual analysis limits by a factor of two in some cases. In the \SU2 singlet (doublet) representation, \VLTa production is entirely excluded for $\kappa > 0.57$ (0.72), with the requirement that $\RW < 50\%$. For $\RW < 5\%$, \SU2 singlet \VLTa production is excluded for $\MT< 1.5$~TeV and $\kappa < 0.25$.

In the benchmark scenario of \SU2 singlet \VLTa production with $\RW = 20\%$, masses less than 2.0~TeV are excluded at the 95\% CL, while in the \SU2 doublet representation with $\RW = 20\%$ masses less than 1.4~TeV are excluded. The highest excluded mass is found for the \SU2 singlet representation assuming $\RW = 50\%$, where masses up to 2.1~TeV are excluded. This significantly surpasses the reach of the individual searches. These results provide the most restrictive limits on the single production of $T$ quarks to date.

% End of text imported from the .//Sections/Conclusions.tex input file

\FloatBarrier
\section*{Acknowledgements}
%

%

% The next lines are included from the .//acknowledgements/Acknowledgements.tex input file
%
%

%
%

We thank CERN for the very successful operation of the LHC and its injectors, as well as the support staff at
CERN and at our institutions worldwide without whom ATLAS could not be operated efficiently.

The crucial computing support from all WLCG partners is acknowledged gratefully, in particular from CERN, the ATLAS Tier-1 facilities at TRIUMF/SFU (Canada), NDGF (Denmark, Norway, Sweden), CC-IN2P3 (France), KIT/GridKA (Germany), INFN-CNAF (Italy), NL-T1 (Netherlands), PIC (Spain), RAL (UK) and BNL (USA), the Tier-2 facilities worldwide and large non-WLCG resource providers. Major contributors of computing resources are listed in Ref.~\cite{ATL-SOFT-PUB-2023-001}.

We gratefully acknowledge the support of ANPCyT, Argentina; YerPhI, Armenia; ARC, Australia; BMWFW and FWF, Austria; ANAS, Azerbaijan; CNPq and FAPESP, Brazil; NSERC, NRC and CFI, Canada; CERN; ANID, Chile; CAS, MOST and NSFC, China; Minciencias, Colombia; MEYS CR, Czech Republic; DNRF and DNSRC, Denmark; IN2P3-CNRS and CEA-DRF/IRFU, France; SRNSFG, Georgia; BMBF, HGF and MPG, Germany; GSRI, Greece; RGC and Hong Kong SAR, China; ISF and Benoziyo Center, Israel; INFN, Italy; MEXT and JSPS, Japan; CNRST, Morocco; NWO, Netherlands; RCN, Norway; MNiSW, Poland; FCT, Portugal; MNE/IFA, Romania; MSTDI, Serbia; MSSR, Slovakia; ARIS and MVZI, Slovenia; DSI/NRF, South Africa; MICIU/AEI, Spain; SRC and Wallenberg Foundation, Sweden; SERI, SNSF and Cantons of Bern and Geneva, Switzerland; NSTC, Taipei; TENMAK, T\"urkiye; STFC/UKRI, United Kingdom; DOE and NSF, United States of America.

Individual groups and members have received support from BCKDF, CANARIE, CRC and DRAC, Canada; CERN-CZ, FORTE and PRIMUS, Czech Republic; COST, ERC, ERDF, Horizon 2020, ICSC-NextGenerationEU and Marie Sk{\l}odowska-Curie Actions, European Union; Investissements d'Avenir Labex, Investissements d'Avenir Idex and ANR, France; DFG and AvH Foundation, Germany; Herakleitos, Thales and Aristeia programmes co-financed by EU-ESF and the Greek NSRF, Greece; BSF-NSF and MINERVA, Israel; NCN and NAWA, Poland; La Caixa Banking Foundation, CERCA Programme Generalitat de Catalunya and PROMETEO and GenT Programmes Generalitat Valenciana, Spain; G\"{o}ran Gustafssons Stiftelse, Sweden; The Royal Society and Leverhulme Trust, United Kingdom.

In addition, individual members wish to acknowledge support from Armenia: Yerevan Physics Institute (FAPERJ); CERN: European Organization for Nuclear Research (CERN PJAS); Chile: Agencia Nacional de Investigaci\'on y Desarrollo (FONDECYT 1230812, FONDECYT 1230987, FONDECYT 1240864); China: Chinese Ministry of Science and Technology (MOST-2023YFA1605700), National Natural Science Foundation of China (NSFC - 12175119, NSFC 12275265, NSFC-12075060); Czech Republic: Czech Science Foundation (GACR - 24-11373S), Ministry of Education Youth and Sports (FORTE CZ.02.01.01/00/22\_008/0004632), PRIMUS Research Programme (PRIMUS/21/SCI/017); EU: H2020 European Research Council (ERC - 101002463); European Union: European Research Council (ERC - 948254, ERC 101089007), Horizon 2020 Framework Programme (MUCCA - CHIST-ERA-19-XAI-00), European Union, Future Artificial Intelligence Research (FAIR-NextGenerationEU PE00000013), Italian Center for High Performance Computing, Big Data and Quantum Computing (ICSC, NextGenerationEU); France: Agence Nationale de la Recherche (ANR-20-CE31-0013, ANR-21-CE31-0013, ANR-21-CE31-0022, ANR-22-EDIR-0002), Investissements d'Avenir Labex (ANR-11-LABX-0012); Germany: Baden-Württemberg Stiftung (BW Stiftung-Postdoc Eliteprogramme), Deutsche Forschungsgemeinschaft (DFG - 469666862, DFG - CR 312/5-2); Italy: Istituto Nazionale di Fisica Nucleare (ICSC, NextGenerationEU), Ministero dell'Università e della Ricerca (PRIN - 20223N7F8K - PNRR M4.C2.1.1); Japan: Japan Society for the Promotion of Science (JSPS KAKENHI JP22H01227, JSPS KAKENHI JP22H04944, JSPS KAKENHI JP22KK0227, JSPS KAKENHI JP23KK0245); Netherlands: Netherlands Organisation for Scientific Research (NWO Veni 2020 - VI.Veni.202.179); Norway: Research Council of Norway (RCN-314472); Poland: Ministry of Science and Higher Education (IDUB AGH, POB8, D4 no 9722), Polish National Agency for Academic Exchange (PPN/PPO/2020/1/00002/U/00001), Polish National Science Centre (NCN 2021/42/E/ST2/00350, NCN OPUS nr 2022/47/B/ST2/03059, NCN UMO-2019/34/E/ST2/00393, NCN \& H2020 MSCA 945339, UMO-2020/37/B/ST2/01043, UMO-2021/40/C/ST2/00187, UMO-2022/47/O/ST2/00148, UMO-2023/49/B/ST2/04085, UMO-2023/51/B/ST2/00920); Slovenia: Slovenian Research Agency (ARIS grant J1-3010); Spain: Generalitat Valenciana (Artemisa, FEDER, IDIFEDER/2018/048), Ministry of Science and Innovation (MCIN \& NextGenEU PCI2022-135018-2, MICIN \& FEDER PID2021-125273NB, RYC2019-028510-I, RYC2020-030254-I, RYC2021-031273-I, RYC2022-038164-I), PROMETEO and GenT Programmes Generalitat Valenciana (CIDEGENT/2019/027); Sweden: Carl Trygger Foundation (Carl Trygger Foundation CTS 22:2312), Swedish Research Council (Swedish Research Council 2023-04654, VR 2018-00482, VR 2022-03845, VR 2022-04683, VR 2023-03403, VR grant 2021-03651), Knut and Alice Wallenberg Foundation (KAW 2018.0157, KAW 2018.0458, KAW 2019.0447, KAW 2022.0358); Switzerland: Swiss National Science Foundation (SNSF - PCEFP2\_194658); United Kingdom: Leverhulme Trust (Leverhulme Trust RPG-2020-004), Royal Society (NIF-R1-231091); United States of America: U.S. Department of Energy (ECA DE-AC02-76SF00515), Neubauer Family Foundation.

%
%

% End of text imported from the .//acknowledgements/Acknowledgements.tex input file

%
%
%
%
%
%
%
%
%
%

%
%

%
%
%
\printbibliography
\clearpage
% ATLAS Collaboration author list
% Reference date of EXOT-2021-02 is 2024-06-20
% Author list last updated on date 12-NOV-24
% Data extracted on 12-Nov-2024 for paper reference EXOT-2021-02
% at 11:53am
 
\begin{flushleft}
\hypersetup{urlcolor=black}
{\Large The ATLAS Collaboration}

\bigskip

\AtlasOrcid[0000-0002-6665-4934]{G.~Aad}$^\textrm{\scriptsize 104}$,
\AtlasOrcid[0000-0001-7616-1554]{E.~Aakvaag}$^\textrm{\scriptsize 17}$,
\AtlasOrcid[0000-0002-5888-2734]{B.~Abbott}$^\textrm{\scriptsize 123}$,
\AtlasOrcid[0000-0002-0287-5869]{S.~Abdelhameed}$^\textrm{\scriptsize 119a}$,
\AtlasOrcid[0000-0002-1002-1652]{K.~Abeling}$^\textrm{\scriptsize 56}$,
\AtlasOrcid[0000-0001-5763-2760]{N.J.~Abicht}$^\textrm{\scriptsize 50}$,
\AtlasOrcid[0000-0002-8496-9294]{S.H.~Abidi}$^\textrm{\scriptsize 30}$,
\AtlasOrcid[0009-0003-6578-220X]{M.~Aboelela}$^\textrm{\scriptsize 45}$,
\AtlasOrcid[0000-0002-9987-2292]{A.~Aboulhorma}$^\textrm{\scriptsize 36e}$,
\AtlasOrcid[0000-0001-5329-6640]{H.~Abramowicz}$^\textrm{\scriptsize 155}$
% ...
\bigskip
\\

% ...
$^{17}$Department for Physics and Technology, University of Bergen, Bergen; Norway.\\
% ...
$^{30}$Physics Department, Brookhaven National Laboratory, Upton NY; United States of America.\\
% ...
$^{36}$$^{(a)}$Facult\'e des Sciences Ain Chock, Universit\'e Hassan II de Casablanca;$^{(b)}$Facult\'{e} des Sciences, Universit\'{e} Ibn-Tofail, K\'{e}nitra;$^{(c)}$Facult\'e des Sciences Semlalia, Universit\'e Cadi Ayyad, LPHEA-Marrakech;$^{(d)}$LPMR, Facult\'e des Sciences, Universit\'e Mohamed Premier, Oujda;$^{(e)}$Facult\'e des sciences, Universit\'e Mohammed V, Rabat;$^{(f)}$Institute of Applied Physics, Mohammed VI Polytechnic University, Ben Guerir; Morocco.\\
% ...
$^{45}$Physics Department, Southern Methodist University, Dallas TX; United States of America.\\
% ...
$^{50}$Fakult\"{a}t Physik , Technische Universit{\"a}t Dortmund, Dortmund; Germany.\\
% ...
$^{56}$II. Physikalisches Institut, Georg-August-Universit\"{a}t G\"ottingen, G\"ottingen; Germany.\\
% ...
$^{104}$CPPM, Aix-Marseille Universit\'e, CNRS/IN2P3, Marseille; France.\\
% ...
$^{119}$$^{(a)}$New York University Abu Dhabi, Abu Dhabi;$^{(b)}$United Arab Emirates University, Al Ain; United Arab Emirates.\\
% ...
$^{123}$Homer L. Dodge Department of Physics and Astronomy, University of Oklahoma, Norman OK; United States of America.\\
% ...
$^{155}$Raymond and Beverly Sackler School of Physics and Astronomy, Tel Aviv University, Tel Aviv; Israel.\\
% ...
$^{a}$ Also Affiliated with an institute covered by a cooperation agreement with CERN.\\
$^{b}$ Also at An-Najah National University, Nablus; Palestine.\\
$^{c}$ Also at Borough of Manhattan Community College, City University of New York, New York NY; United States of America.\\
$^{d}$ Also at Center for Interdisciplinary Research and Innovation (CIRI-AUTH), Thessaloniki; Greece.\\
$^{e}$ Also at CERN, Geneva; Switzerland.\\
$^{f}$ Also at CMD-AC UNEC Research Center, Azerbaijan State University of Economics (UNEC); Azerbaijan.\\
% ...

\end{flushleft}

% Created with Glance <Atlas.Glance@cern.ch>

%

%
%

%

%
%
%
%

\end{document}